\documentclass{iaus}
\usepackage{graphicx}

\def\apj{ApJ}%
\def\apjl{ApJ}%
\def\apjs{ApJS}%
\def\aap{A\&A}%
\def\mnras{MNRAS}%
\def\pasp{PASP}%
\def\pasj{PASJ}%
\def\nat{Nature}%
\title [The Rossiter-McLaughlin effect]{The long history of the Rossiter-McLaughlin effect and its
  recent applications} 

\author[Simon Albrecht]   
{Simon Albrecht$^1$}

\affiliation{$^1$Massachusetts Institute of Technology, Kavli Institute for Astrophysics and Space Research,
  Cambridge, MA 02139, USA}  

\pubyear{2011}
\volume{282}  
\pagerange{100--106}
\jname{From Interacting Binaries to Exoplanets: Essential Modeling Tools}
\editors{A.C. Editor, B.D. Editor \& C.E. Editor, eds.}
\begin{document}

\maketitle

\begin{abstract}
In this paper I will review the Rossiter-McLaughlin (RM) effect; its
  history, how it manifests itself during stellar eclipses and
  planetary transits, and the increasingly important role its
  measurements play in guiding our understanding of the formation and
  evolution of close binary stars and exoplanet systems.

\end{abstract}

\firstsection 
\section{Introduction}

The sun is the only star for which we can obtain detailed information
on spacial scales much smaller than its diameter. For some nearby
stars or giant stars optical/infrared long baseline interferometry
does give information on scales comparable to the stellar size
\cite[(e.g. Baines et al.\ 2010)]{baines2010}. For most stars, however, we are
not able to resolve their surfaces. These stars are essentially
point sources, even for the biggest telescopes.

This is a pity as many questions in stellar astrophysics and astronomy
would benefit from such knowledge. Astronomers have therefore
developed a number of techniques to overcome this limitation. For
example Doppler imaging \cite[(Strassmeier 2002)]{strassmeier2002},
polarimetry (see K. Bjorkman these proceedings), or tomography (see
M. Richards these proceedings) let us gain under certain conditions
information on small spatial features. Close binary
star systems with orbits of only a few days or stars harbouring extra
solar planets (exoplanets) can provide us with an additional
opportunity to obtain high spatial resolution, if the line of sight
lies in the orbital plane. In such cases eclipses or transits may be
observed.

During eclipses or transits telescopes integrate not over the complete
stellar disk, as parts are hidden from view. Comparing the amount of
light obtained at different phases of eclipses with the light received
out of eclipse, system parameters like ratios of the radii of the two
objects, orbital inclination and possible inhomogeneities on the
stellar surface of the background star, like star spots can be
determined (e.g.\ C.\ Maceroni these proceeding, \cite[Huber 2010]{Huber2010}).

What properties can be studied if we are not only to record the amount
of light blocked from view, but also record the dimming as function of
the wavelength? Already 1893 Holt realized that observing an eclipse
with a spectrograph which has a high enough spectral resolution to
resolve stellar absorption lines, will lead to inside knowledge on
stellar rotation \cite[(Holt 1893)]{holt1893}. Stellar lines are
broadened by Doppler shift due to rotation. Light emitted from
approaching stellar surface areas is blue shifted and light emitted
from receding stellar surface areas is red shifted. During eclipse
parts of the rotating stellar surface is hidden, causing a weakening of
the corresponding velocity component of the stellar absorption
lines. Modeling of this spectral distortion reveals the projected
stellar rotation speed ($v \sin i_{\star}$) and the angle between the
stellar and orbital spins projected on the plane of the sky: the
projected obliquity.\footnote{This angle is denoted either $\beta$
  after \cite[Hosokawa (1953)]{hosokawa1953} or $\lambda$ after
  \cite[Ohta et al.\ (2005)]{ohta2005}, $\lambda=-\beta$.}

A claim of the detection of the rotation anomaly was made by
\cite[Schlesinger (1910)]{schlesinger1910}, but more definitive
measurements were achieved by \cite[Rossiter (1924)]{rossiter1924} and
\cite[McLaughlin (1924)]{mclaughlin1924} for the $\beta$\,Lyrae and
Algol systems, respectively. These researchers reported the change of
the first moment of the absorption lines, sometimes called center of
gravity, derived form the shape of the absorption line.  \cite[Struve
\& Elvey (1931)]{struve1931} reported the shape and its change during
eclipse in the Algol system. The phenomenon is now known as the
Rossiter-McLaughlin (RM) effect. Various aspects of the theory of the
effect have been worked out by \cite[ Hosokawa (1953), Kopal (1959),
Sato (1974), Ohta et al.\ (2005), Gimenez (2006), Hadrava (2009),
Hirano et al.\ (2010) and Hirano et al.\ (2011a)]{hosokawa1953,
  kopal1959, sato1974, ohta2005,
  gimenez2006,hadrava2009,hirano2010,hirano2011a}.

\section{The RM effect and some quantities which can be measured
  with it}

\cite[Holt (1893)]{holt1893} realised that the rotation anomaly,
occuring during eclipses, is a opportunity to measure $v \sin
i_{\star}$ independently from a measurement of the width of absorption
lines. Measuring $v \sin i_{\star}$ from line widths is challenging as
these are influenced not only by rotation but also other processes,
most notably by velocity fields on the stellar surface and pressure
broadening. The strengths of these mechanisms are often not precisely
known, introducing a substantial uncertainty in the $v \sin i_{\star}$
measurement even if the width of the line can be determined with high
accuracy \cite[(e.g Valenti \& Fischer 2005)]{valenti2005}. The
amplitude of the RM effect is not as strongly influenced by these
broadening mechanisms, making it an interesting tool for measuring $v
\sin i_{\star}$ in particular cases \cite[(e.g. Twigg 1979, Worek et
al.\ 1988, Rucinski et al.\ 2009)]{twigg1979,worek1988,rucinski2009}.
In addition, if differential rotation is present then it might be
detected in fortunate cases via the RM effect \cite[(Hosokawa 1953,
Hirano et al.\ 2011a)]{hosokawa1953,hirano2011a}.  Currently, however,
the RM effect is mainly seen as a tool to obtain the projection of
stellar obliquity, an observable hard or impossible to measure
otherwise.

However, not only stellar rotation can be studied. With the help of the
differential RM effect atmospheres of transiting planets may be
studied \cite[(Snellen 2004, Dreizler et al.\
2009)]{snellen2004,dreizler2009}. The RM effect might also aid in the
search and confirmation of planet candidates \cite[(Gaudi \& Winn
2007)]{gaudi2007} or even exomoons \cite[(Simon et al.\
2010)]{simon2010}. Also accretion in an interacting binary might be
studied via the RM effect \cite[(e.g.\ Lehmann \& Mkrtichian
2004)]{lehmann2004}.

\section{The RM effect and obliquities in extrasolar planetary systems}

The properties of exoplanets discovered over the last years have been
very surprising. Many exoplanets orbit their hosts stars on eccentric
orbits and giant planets have been found on orbits with periods of
only a few days ('Hot-Jupiters'). These findings present challenges for
planet formation theories as it is thought that giant planets can only
form at distances of several AU from their host stars, where the
radiation is less harsh and small particles can survive long enough to
build a rocky core which attracts the gaseous envelope from the disk.

Different classes of migration processes have been proposed which
might transport giant planets from their presumed birthplaces inward
to a fraction of an astronomical unit where we find them. Some of
these processes are expected to change the relative orientation
between the stellar and orbital spin \cite[(e.g.\ Nagasawa et al.\ 2008,
Fabrycky \& Tremaine 2007)]{nagasawa2008,fabrycky2007}, while others
will conserve the relative orientation \cite[(Lin et al.\
1996)]{lin1996}, or even reduce a possible misalignment
\cite[(Cresswell et al.\ 2007)]{cresswell2007}. Therefore measuring
the obliquity of these systems will lead to inside knowledge of the
formation and evolution of these systems.
 
\begin{figure}[t]
\begin{center}
\includegraphics[width=9.cm]{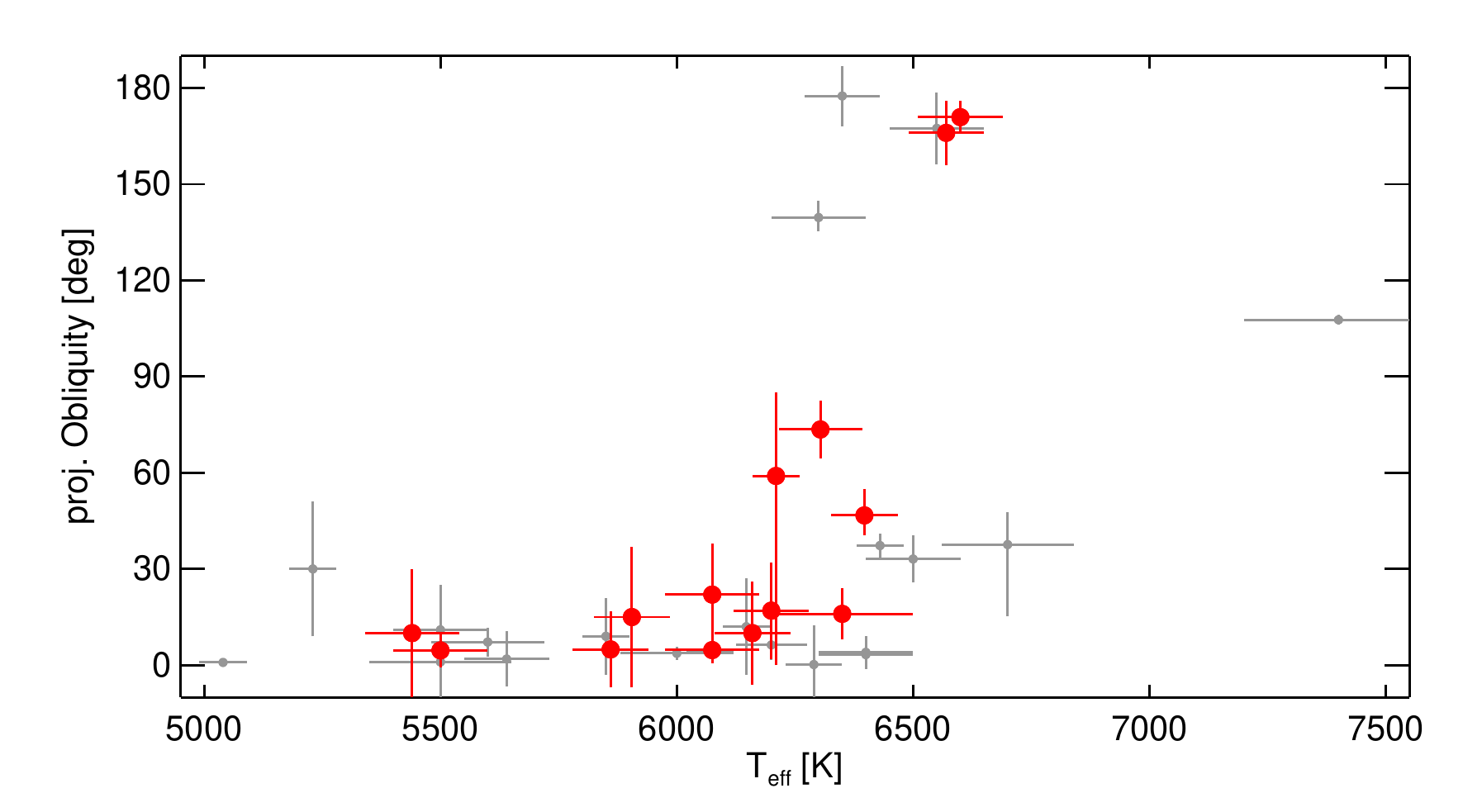} 
\caption{{\bf Hotter stars have oblique rotation.} The projected
  obliquity of Hot Jupiter (M$_{planet} >0.2 $ M$_{\rm Jupiter}$;
  Period $< 6$ days) systems is plotted as function of the effective
  temperature of the host star. Using measurements available at that time,
  \cite[Winn et al.\,2010a]{winn2010} noticed that systems with cool
  stars are aligned, while the obliquities of hot stars tends to be
  higher (gray small circles). Since then 16 new RM measurements have
  been reported (red large circles). The systems Kepler-8, CoRoT-1/11,
  have been omitted (see section~\ref{challenges}) and the values for
  WASP-1/2 have been taken from \cite[Albrecht et al.\
  (2011)]{albrecht2011b}}
   \label{fig1}
\end{center}
\end{figure}

\subsection{Results of RM measurements}

The first measurement of a projected obliquity in an extrasolar
system was made by \cite[Queloz et al.\ (2000)]{queloz2000}. They
found that HD\,209458 has a low obliquity. Over the following years
the angle between the stellar and orbital spins have been measured in
about 30 systems. It was found that for some of these systems the
orbits are inclined or even retrograde with respect to the rotational
spins of their host stars \cite[(see e.g. H\'ebrard et al.\ 2008, Winn
et al.\ 2009, Triaud et al.\ 2010, Simpson et al.\,2011)]{hebrard2008,
  winn2009,triaud2010,simpson2011}.  \cite[Winn et al.\
(2010a)]{winn2010} found that close-in giant planets tend to have
orbits aligned with the stellar spin if the effective temperature
($T_{\rm eff}$) of their host star is $\lesssim 6250$\,K and misaligned
otherwise. \cite[Schlaufman (2010)]{schlaufman2010} obtained similar
results measuring the inclination of spin axes along the line of
sight. \cite[Winn et al.\ (2010a)]{winn2010} further speculated that
this might indicate that {\it all} giant planets are transported
inward by processes which randomize the obliquity. In this picture
tidal waves raised on the star by the close in planet realign the two
angular momentum vectors. The realignment time scale would be short
for planets around stars with convective envelopes ($T_{\rm eff}
\lesssim 6250$\,K), but long, compared to the lifetime of the system, if
the star does not have a convective envelope ($T_{\rm eff} \gtrsim
6250$\,K). Over the last year the RM effect was measured in another 16
systems, and the predictions made by \cite[Winn et al.\
(2010a)]{winn2010} have been confirmed for these systems (see
Fig~\ref{fig1})\footnote{ Rene Heller maintains a webpage with updated
  information of obliquity measurements: {\tt
    http://www.aip.de/People/rheller/content/main\_spinorbit.html} }.

\begin{figure}[t]
\begin{center}
\includegraphics[width=8.8cm]{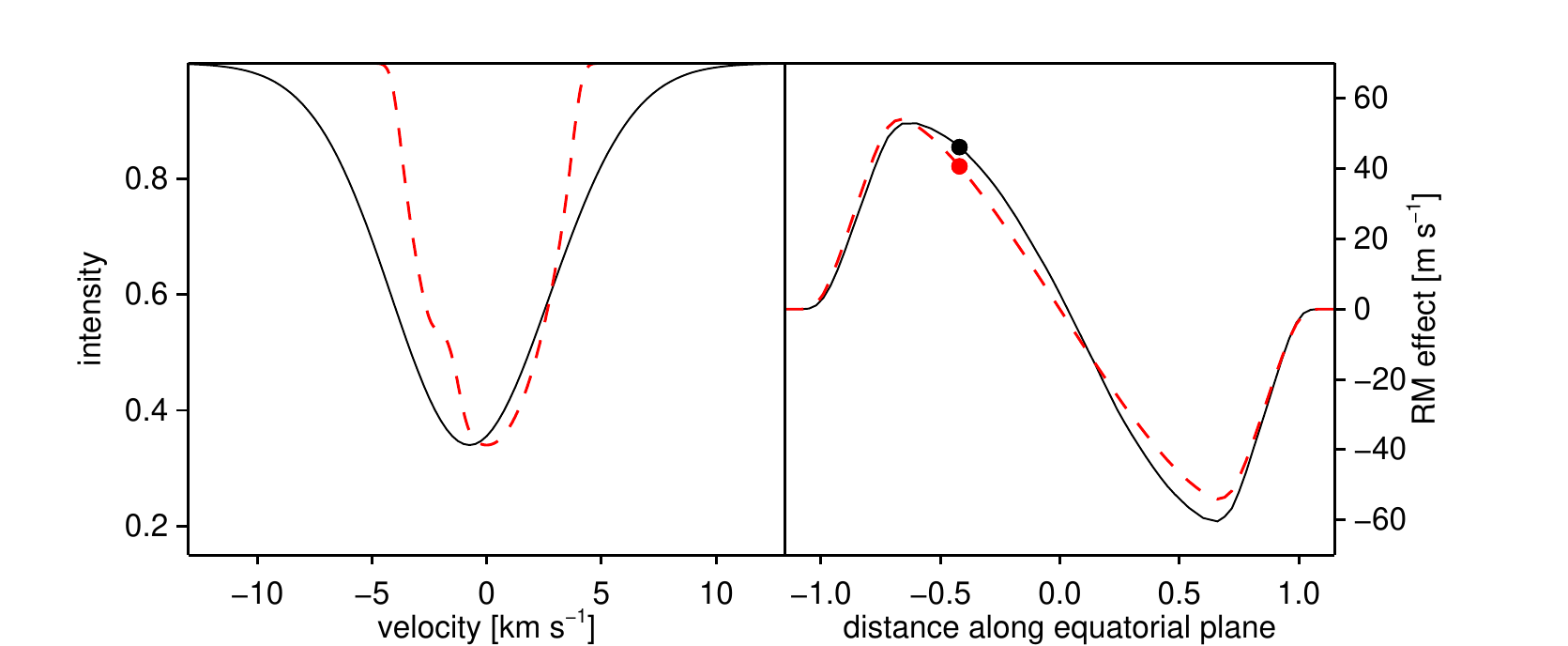} 
\caption{{\bf Line broadening mechanisms and their effect on the RM
    signal.} The left panel shows a model of a absorption line
  broadened by solid body rotation only (red) dashed line and a model
  of a line taking also macro turbulence, convective blue shift and
  solar like differential rotation in account. The right panel shows
  the RM effect for both models. The circles indicate the transit
  phase when the snapshots of the absorption lines on the left side
  have been taken. On can see how the lines as well as the expected RM
  effect differ.}
   \label{fig2}
\end{center}
\end{figure}

\subsection{Challenges}
\label{challenges}

When analysing RM measurements there are challenges which need to be
overcome before a robust estimation of the stellar spin can be
derived.  Not only stellar rotation effects the meaured stellar
absorption lines. They are also broadened by stellar rotation fields
and the point spread function of the spectrograph. Lines are also not
strictly symmetric due to the convective blue shift \cite[(Shporer \&
Brown 2011)]{shporer2011}. See Fig.~\ref{fig2} for an illustration of
this effect. In addition not the center of line is measured (the
quantity most often used by descriptions of the RM effect), but a
cross correlation between a template and the recoded spectrum during
transit (\cite[Hirano et al.\ 2011a)]{hirano2011a}. For the measurement
process additional complications can arise.

\begin{itemize}
\item Similar to transit photometry observations before and after
  transit are important. The RM effect needs to be isolated from other
  sources of RV variations (orbital movements star spots, unknown
  companions,..).  We therefore suspect the uncertainty in the
  Kepler-8 system to be greater then reported by \cite[Jenkins et al.\
  (2010)]{jenkins2010}.
\item Analyzing low SNR RV data can lead to results which are
  systematically biased. This was the case for WASP-2 for which a
  retrograde orbit was reported by \cite[Triaud et al.\
  (2010)]{triaud2010}, but it was later found that from the currently
  available data no information on the obliquity can be derived. See
  \cite[Albrecht et al.\ (2011b)]{albrecht2011b} for details.
\item For systems nearly edge-on (i.e. low impact systems) there
  exists a strong degeneracy between $v \sin i_{\star}$ and the
  projected obliquity and care has to be taken when applying
  photometric and spectroscopic priors. This is the case for WASP-1
  \cite[(Simpson et al.\ 2011, Albrecht et al.\
  2011b)]{simpson2011,albrecht2011b}.
\end{itemize}

\section{Eclipsing binaries}

Although it has been more than $80$ years since the first RM
measurements in binaries, there are relatively few quantitative
analyses of the RM effect in these systems. In the past, observing the
RM effect was generally either avoided (as a hindrance to measuring
accurate spectroscopic orbits) or used to estimate stellar rotation
speeds. Almost all authors explicitly or implicitly assumed that the
orbital and stellar spins were aligned. This lack of measurements is a
pity as the knowledge of obliquity might guide our understanding of
binary formation, in particular the formation of close binaries
(e.g. Fabrycky \& Tremaine 2007, Albrecht et al.\ 2011a).

There is a complication in the RM measurement relative to the low-mass
companion or exoplanet case, if one wants to measure the RM effect in
double lined binaries.  Also the foreground object emits light and
contributes to the observed spectrum. Measuring the center of gravity
of absorption lines would lead to erroneous results.  \cite[Albrecht
et al.\ (2007)]{albrecht2007} therefore developed a method to model
the stellar absorption lines during occultations. A similar method was
also employed in exoplanet systems \cite[(Collier Cameron et al.\
2010)]{cameron2010}.

The BANANA project (see Albrecht et al.\ these proceedings) aims to
measure the projected obliquities in a number of eclipsing binaries to
understand what sets systems with spin-orbit alignment apart from
systems where the spins are not aligned. They find that alignment is
not a simple function of orbital separation or eccentricity.

Another project led by Amaury Triaud aims to measure obliquities in
binaries with F star primaries and late type secondaries.  (A. Triaud
these proceedings).

\section{Outlook}

The future for RM-measurements looks bright. Not only the number of
known eclipsing binaries and transiting exoplanets will increase
thanks to missions like {\it Kepler}, but these missions will also
discover long period systems and systems with multiple transiting
planets. Also the obliquities in systems with smaller planets, likely
to have a different formation history, can be probed \cite[(Winn et
al.\ 2010b, Hirano et al.\ 2011b)]{winn2010c,hirano2011b}. With an
improved understanding of the RM effect we might also be able to
measure in a few systems some second order effects, as described
above.

Stellar obliquities will also be measured by other techniques, like
the method employed by \cite[Schlaufman (2010)]{schlaufman2010}, which
is not as accurate as RM measurements, but has the virtue that it does
not require transit observations. For slowly rotating stars the
crossing of star-spots can be used as tracer of stellar obliquity
\cite[(e.g. Sanchis-Ojeda \& Winn 2011)]{sanchis2011}. For fast
rotating stars which exhibit gravity darkening the projected obliquity
can be estimated from high quality photometry \cite[(Szabo et al.\
2011)]{szabo2011}. Having very precise photometry further opens the
possibility to measure obliquities via the photometric RM effect
\cite[(Groot 2011, Shpoorer et al.\
2011)]{groot2011,shpoorer2011b}. Finally optical interferometry is now
able to measure the projected obliquity for some nearby systems
\cite[(Le Bouquin et al.\ 2009)]{lebouquin2009}. Therefore there is
the chance that our understanding of stellar obliquity, so far an
elusive quantity, will be greatly improved over the coming years.\\\\

\noindent {\it Acknowledgements} I am grateful to the scientific and
local organising committees for organizing a very interesting and
stimulating meeting. I am grateful to Josh Winn for valuable comments
on the manuscript. I am thankful to Dan Fabrycky, Teru Hirano, John
Johnson, Roberto Sanchis-Ojeda, Andreas Quirrenbach, Sabine Reffert,
Johny Setiawan, Ignas Snellen, Willie Torres and Josh Winn for
interesting discussions on stellar obliquities.

\begin{discussion}

\discuss{Piercarlo Bonifacio}{Since you translated effective
  temperatures to masses in your Teff-Obliquity relation, I assume all
  your stars are dwarfs. If the physical parameter determining the
  trend is really mass, you should be able to find some cool massive
  giants with a high obliquity planet.}

\discuss{Simon Albrecht}{That is correct. We only have RM measurements for dwarf stars. Unfortunately,
  it is very difficult to detect transiting planets around giants as
  the radius ratio is so big. Also the R-M measurements would be very
  difficult.}

\discuss{QUESTION}{Do you have any bias in your sample of exoplanets? }

\discuss{Simon Albrecht}{Yes we inherit for example the biases from the planet search
surveys. }

\end{discussion}


\begin{thebibliography}{48}
\expandafter\ifx\csname natexlab\endcsname\relax\def\natexlab#1{#1}\fi

\bibitem[{{Albrecht} {et~al. }(2007)}]{albrecht2007}
{Albrecht}, S., {Reffert}, S., {Snellen}, I., {Quirrenbach}, A., \& {Mitchell},
  D.~S. 2007, \aap, 474, 565

\bibitem[{{Albrecht} {et~al. }(2011)}]{albrecht2011a}
{Albrecht}, S., {Winn}, J.~N., {Carter}, J.~A., {Snellen}, I.~A.~G., et al. 2011{\natexlab{a}}, \apj, 726, 68

\bibitem[{{Albrecht} {et~al.}(2011)}]{albrecht2011b}
{Albrecht}, S., {Winn}, J.~N., {Johnson}, J.~A., {Butler}, et al. 2011{\natexlab{b}}, \apj, 738, 50

\bibitem[{{Baines} {et~al.}(2010)}]{baines2010}
{Baines}, E.~K., {D{\"o}llinger}, M.~P., {Cusano}, F.,	{Guenther},
E.~W. et al. 2010 \apj, 710, 1365

\bibitem[{{Collier Cameron} {et~al.}(2010)}]{cameron2010}
{Collier Cameron}, A., {Bruce}, V.~A., {Miller}, G.~R.~M., {Triaud},
  A.~H.~M.~J.,  et al. 2010, \mnras, 403, 151

\bibitem[{{Cresswell} {et~al.}(2007)}]{cresswell2007}
{Cresswell}, P., {Dirksen}, G., {Kley}, W., \& {Nelson}, R.~P. 2007, \aap, 473,
  329

\bibitem[{{Dreizler} {et~al.}(2009)}]{dreizler2009}
{Dreizler}, S., {Reiners}, A., {Homeier}, D., \& {Noll}, M. 2009, \aap, 499,
  615

\bibitem[{{Fabrycky} \& {Tremaine}(2007)}]{fabrycky2007}
{Fabrycky}, D. \& {Tremaine}, S. 2007, \apj, 669, 1298

\bibitem[{{Gaudi} \& {Winn}(2007)}]{gaudi2007}
{Gaudi}, B.~S. \& {Winn}, J.~N. 2007, \apj, 655, 550

\bibitem[{{Gim{\'e}nez}(2006)}]{gimenez2006}
{Gim{\'e}nez}, A. 2006, \apj, 650, 408

\bibitem[{{Groot}(2011)}]{groot2011}
{Groot}, P.~J. 2011, ArXiv:1104.3428

\bibitem[{{Hadrava}(2009)}]{hadrava2009}
{Hadrava}, P. 2009, ArXiv:0909.0172

\bibitem[{{H{\'e}brard} {et~al.}(2008)}]{hebrard2008}
{H{\'e}brard}, G., {et~al.} 2008, \aap, 488, 763

\bibitem[{{Hirano} {et~al.}(2011a)}]{hirano2011a}
{Hirano}, T., {Suto}, Y., {Winn}, J.~N., {Taruya}, A.,et al. 2011, arXiv:1108.4430

\bibitem[{{Hirano} {et~al.}(2011b)}]{hirano2011b}
{Hirano}, T., {Narita}, N., {Shporer}, A., {Sato}, B.,et al. 2011, \pasj, 63, 531

\bibitem[{{Hirano} {et~al.}(2010)}]{hirano2010}
{Hirano}, T., {Suto}, Y., {Taruya}, A., {Narita}, N., et al. 2010, \apj, 709, 458

\bibitem[{{Holt}(1893)}]{holt1893}
{Holt}, J.~R. 1893, \aap, 12, 646

\bibitem[{{Huber}{et~al.}(2010)}]{huber2010} 
{Huber}, K.~F., {Czesla}, S., {Wolter}, U., {Schmitt},  J.~H.~M.~M \aap, 514, A39

\bibitem[{{Hirano} {et~al.}(2010)}]{hirano2010}
{Hirano}, T., {Suto}, Y., {Taruya}, A., {Narita}, N., et al. \apj, 709, 458

\bibitem[{{Jenkins} {et~al.}(2010)}]{jenkins2010}
{Jenkins}, J.~M., {Borucki}, W.~J., {Koch}, D.~G., {Marcy}, et al.  2010, \apj, 724, 1108

\bibitem[{{Kopal}(1959)}]{kopal1959}
{Kopal}, Z. 1959, {Close binary systems} (The International Astrophysics
  Series, London: Chapman \& Hall, 1959)

\bibitem[{{Le Bouquin} {et~al.}(2009)}]{lebouquin2009}
{Le Bouquin}, J., {Absil}, O., {Benisty}, M., {Massi}, F., et al. 2009, \aap, 498, L41

\bibitem[{{Lin} {et~al.}(1996)}]{lin1996}
{Lin}, D.~N.~C., {Bodenheimer}, P., \& {Richardson}, D.~C. 1996, \nat, 380, 606

\bibitem[{{Lehmann} \& {Mkrtichian}(2004)}]{lehmann2004}
{Lehmann}, H. \& {Mkrtichian}, D.~E. 2004, \aap, 413, 293

\bibitem[{{McLaughlin}(1924)}]{mclaughlin1924}
{McLaughlin}, D.~B. 1924, \apj, 60, 22

\bibitem[{{Nagasawa} {et~al.}(2008)}]{nagasawa2008}
{Nagasawa}, M., {Ida}, S., \& {Bessho}, T. 2008, \apj, 678, 498

\bibitem[{{Ohta} {et~al.}(2005)}]{ohta2005}
{Ohta}, Y., {Taruya}, A., \& {Suto}, Y. 2005, \apj, 622, 1118

\bibitem[{{Queloz} {et~al.}(2000)}]{queloz2000}
{Queloz}, D., {Eggenberger}, A., {Mayor}, M., {Perrier}, C., et al. 2000, \aap, 359, L13

\bibitem[{{Rossiter}(1924)}]{rossiter1924}
{Rossiter}, R.~A. 1924, \apj, 60, 15

\bibitem[{{Rucinski}(2009)}]{rucinski2009}
{Rucinski}, S.~M. 2009, \mnras, 395, 2299

\bibitem[{{Sanchis-Ojeda} \& {Winn}(2011)}]{sanchis2011}
{Sanchis-Ojeda}, R. \& {Winn}, J.~N. 2011, arXiv:1107.2920

\bibitem[{{Sato}(1974)}]{sato1974}
{Sato}, K. 1974, \pasj, 26, 65

\bibitem[{{Schlaufman}(2010)}]{schlaufman2010}
{Schlaufman}, K.~C. 2010, \apj, 719, 602

\bibitem[{{Schlesinger}(1910)}]{schlesinger1910}
{Schlesinger}, F. 1910, Pub. of the Allegheny Observatory of the
  University of Pittsburgh, 1, 123

\bibitem[{{Shporer} \& {Brown}(2011)}]{shporer2011}
{Shporer}, A. \& {Brown}, T. 2011, \apj, 733, 30

\bibitem[{{Shporer} {et~al.}(2011)}]{shpoorer2011b}
{Shporer}, A., {Brown}, T., {Mazeh}, T., \& {Zucker}, S. 2011, ArXiv:1107.4458

\bibitem[{{Simon} {et~al.}(2010)}]{simon2010}
{Simon}, A.~E., {Szab{\'o}}, G.~M., {Szatm{\'a}ry}, K., \& {Kiss}, L.~L. 2010,
  ArXiv:1004.1143

\bibitem[{{Simpson} {et~al.}(2011)}]{simpson2011}
{Simpson}, E.~K., {Pollacco}, D., {Cameron}, A.~C., {H{\'e}brard}, G.,
et al. 2011, \mnras, 414, 3023

\bibitem[{{Snellen}(2004)}]{snellen2004}
{Snellen}, I.~A.~G. 2004, \mnras, 353, L1

\bibitem[{{Strassmeier}(2002)}]{strassmeier2002}
{Strassmeier}, K.~G. 2002, Astronomische Nachrichten, 323, 309

\bibitem[{{Struve} \& {Elvey}(1931)}]{struve1931}
{Struve}, O. \& {Elvey}, C.~T. 1931, \mnras, 91, 663

\bibitem[{{Szab{\'o}} {et~al.}(2011)}]{szabo2011}
{Szab{\'o}}, G.~M., {Szab{\'o}}, R., {Benk{\H o}}, J.~M., {Lehmann}, H.,
   et al. 2011, \apjl, 736, L4

\bibitem[{{Triaud} {et~al.}(2010)}]{triaud2010}
{Triaud}, A.~H.~M.~J., {Collier Cameron}, A., {Queloz}, D.,
{Anderson}, D.~R., et al. 2010, \aap, 524, 25

\bibitem[{{Twigg}(1979)}]{twigg1979}
{Twigg}, L.~W. 1979, PhD thesis, Florida Univ., Gainesville.

\bibitem[{{Valenti} \& {Fischer}(2005)}]{valenti2005}
{Valenti}, J.~A. \& {Fischer}, D.~A. 2005, \apjs, 159, 141

\bibitem[{{Winn} {et~al.}(2010)}]{winn2010}
{Winn}, J.~N., {Fabrycky}, D., {Albrecht}, S., \& {Johnson}, J.~A.
  2010{\natexlab{a}}, \apjl, 718, L145

\bibitem[{{Winn} {et~al.}(2010)}]{winn2010c}
{Winn}, J.~N., {Johnson}, J.~A., {Howard}, A.~W., {Marcy}, G.~W., et al.
  2010{\natexlab{b}}, \apjl, 723, L223

\bibitem[{{Winn} {et~al.}(2009)}]{winn2009}
{Winn}, J.~N., {Johnson}, J.~A., {Albrecht}, S., {Howard}, A.~W. et al. 2009, \apjl, 703, L99

\bibitem[{{Worek} {et~al.}(1988)}]{worek1988}
{Worek}, T.~F., {Zizka}, E.~R., {King}, M.~W., \& {Kiewiet de Jonge}, J.~H.
  1988, \pasp, 100, 371

\end{thebibliography}
\end{document}